\documentclass[twocolumn,prc,aps, nofootinbib]{revtex4}
\usepackage{graphicx}
\usepackage[section]{placeins}

 \def\gsim{\mathrel{\rlap{\lower4pt\hbox{\hskip1pt$\sim$}}
 \raise1pt\hbox{$>$}}}

 \newcommand\la{\langle}
 \newcommand\ra{\rangle}
 \newcommand\beq{\begin{equation}}
 
 \newcommand\eeq{\end{equation}}
 \newcommand\beqn{\begin{eqnarray}}
 \newcommand\eeqn{\end{eqnarray}}
 \newcommand\eq{\!=\!}
 \newcommand\eqq{\!\equiv\!}
\def\mb{\,\mbox{mb}}
\def\fm{\,\mbox{fm}}
\def\GeV{\,\mbox{GeV}}
\def\TeV{\,\mbox{TeV}}
\def\lsim{\mathrel{\rlap{\lower4pt\hbox{\hskip1pt$\sim$}}
    \raise1pt\hbox{$<$}}}         
\def\gsim{\mathrel{\rlap{\lower4pt\hbox{\hskip1pt$\sim$}}
    \raise1pt\hbox{$>$}}}         

\def\Im{\,\mbox{Im}\,}
\def\mb{\,\mbox{mb}}
\def\fm{\,\mbox{fm}}
\def\GeV{\,\mbox{GeV}}

\def\beq{\begin{equation}}
\def\eeq{\end{equation}}

\def\beqy{\begin{eqnarray}}
\def\eeqy{\end{eqnarray}}

\begin{document}

\title{\bf Quenching of high-\boldmath$p_T$ hadrons:
Energy Loss vs Color Transparency}

\author{B. Z. Kopeliovich$^1$}
\author{J. Nemchik$^{2,3}$}
\author{ I. K. Potashnikova$^1$}
\author{Iv\'an Schmidt$^1$}
\affiliation{\centerline{$^1$Departamento de F\'{\i}sica,
Universidad T\'ecnica Federico Santa Mar\'{\i}a; and}
Instituto de estudios avanzados en Ciencias e Ingenier'a; and\\
Centro Cient\'ifico-Tecnol\'ogico de Valpara\'iso;\\
Casilla 110-V, Valpara\'iso, Chile
\\
{$^{2}$\sl 
Czech Technical University in Prague,
FNSPE, B\v rehov\'a 7,
11519 Prague, Czech Republic
}\\
{$^{3}$\sl
Institute of Experimental Physics SAS, Watsonova 47,
04001 Kosice, Slovakia
}}

\begin{abstract}
High-$p_T$ hadrons produced in hard collisions and detected inclusively
bear peculiar features: (i) they originate from jets whose initial virtuality and energy are of the same order; (ii) such jets are rare and have a very biased energy sharing among the particles, namely, the detected hadron carries the main fraction of the jet energy.
The former feature leads to an extremely intensive gluon radiation and energy dissipation at the early stage of hadronization, either in vacuum or in a medium. As a result, a leading hadron must be produced on a short length scale. Evaluation within a model of perturbative fragmentation confirms the shortness of the production length. This result is at variance
with the unjustified assumption of long production length, made within the popular energy loss scenario.
Thus we conclude that the main reason of suppression of high-$p_T$ hadrons in heavy ion collisions is the controlled by color transparency attenuation of a high-$p_T$ dipole propagating through the hot medium. Adjusting a single parameter, the transport coefficient, we describe quite well the data from LHC and RHIC for the suppression
factor $R_{AA}$ as function of $p_T$, collision energy and centrality. We observe that 
the complementary effect of initial state interaction causes a flattening and even fall of $R_{AA}$ at large $p_T$.  The azimuthal anisotropy of hadron production, calculated with no further adjustment, also agrees well with data at  different energies and centralities.

\end{abstract}



\maketitle

\section{Introduction}

A colored parton produced with a high momentum in a hard reaction hadronizes, 
forming a jet of hadrons. It is natural to expect that the production time of such a jet, averaged over jet configurations, rises with the jet energy due to the effect of Lorentz time dilation. 
Although the jet is detected at macroscopic distances from the collision point, 
its space-time development at the early stages of hadronization can be probed with nuclear targets \cite{within}.  

In this paper we concentrate on the rare type of jets in which the main fraction of the jet momentum is carried by a single (leading) hadron. In some cases, like in semi-inclusive deep-inelastic scattering (SIDIS), such events can be selected explicitly, because the fractional light-cone momentum $z_h$ of the detected hadron can be measured. In high-$p_T$ single hadron production processes  the fractional hadron momentum is unobserved, but the convolution of the steeply falling jet momentum distribution with the fragmentation function picks up high values of $z_h$ (see Sect.~\ref{mean-zh}). Thus, inclusive high-$p_T$ hadron production without observation of the whole jet implicitly selects an unusual type of jets with a very biased sharing of energy. 

Another generic feature of such jets is an extremely high initial virtuality, which is of the same order as the jet energy. This leads to a very  intensive gluon radiation and energy dissipation at the early stage of hadronization. In order to respect energy conservation in the production of a high-$z_h$ hadron, the radiative dissipation of energy must be stopped by the production of a colorless hadronic configuration (QCD dipole or pre-hadron)
on a short time or length (we use both terms interchangeably)  scale. This is considered in detail in Sect.~\ref{lp}, where the rate of radiative energy loss  in vacuum is calculated as a function of time. The production length $l_p$ of a colorless dipole finalizing hadronization is calculated within a model of perturbative hadronization and found to be rather short. The important observation is a weak dependence of $l_p$ on $p_T$, which might look counterintuitive, because the Lorentz factor is expected to stretch $l_p$ at higher $p_T$. However, the rate of energy dissipation increases as well, and this works in the opposite direction, trying to shorten $l_p$.

Since a colorless dipole is created at a short time scale inside a dense medium, it has to survive through the medium in order to be detected. The evolution of the dipole in the medium and its attenuation is the subject of Sect.~\ref{dipole}.
The key phenomenon controlling the dipole surviving probability is color transparency, which corresponds to the enhanced transparency of the medium for small-size dipoles \cite{zkl}.
We employ the relation between the dipole cross section and transport coefficient (broadening) found in \cite{jkt,broadening,Domdey}. Correspondingly, the observed magnitude of hadron attenuation can be used as a probe for the transport coefficient, which characterizes the medium density.

In Sect.~\ref{data} we compare the calculated suppression factor $R_{AA}$ with data, as function of $p_T$, collision energy and centrality of the collision.  This comparison involves only one fitted parameter, $\hat q_0$, which is the maximal transport coefficient of the medium
created in a central collision of given nuclei, at a given energy. Otherwise, this parameter is universal for all observables. The shape of the $p_T$ dependence of $R_{AA}$ is found to be in a good accord with data. In particular, the observed rise of $R_{AA}(p_T)$ at LHC
is easily and naturally explained by the color transparency effect, calculated within the rigorous quantum-mechanical description known as the path-integral technique. 

Comparison with data results in the transport coefficient ,which ranges  from $\hat q_0=1.2\GeV^2\!/\!\fm$ at $\sqrt{s}=62\GeV$ up to $2\GeV^2\!/\!\fm$ at $\sqrt{s}=2.76\TeV$, for collisions of heavy nuclei, gold and lead. These values of the transport coefficient are about twice as large as those that were found in \cite{my-alice} within a simplified model of dipole evolution. Nevertheless, they are an order of magnitude smaller than
what was found in the analysis \cite{phenix-theor}, based on the energy loss scenario
(see e.g. \cite{miklos}), which relies on the unjustified assumption of a long production length $l_p$.

It is worth emphasizing that our approach, based on perturbative QCD, is irrelevant to data at $p_T\lesssim 6\GeV$, which are apparently dominated by hydrodynamics. 

An additional effect related to initial state interactions (ISI) of the colliding nuclei is described in Sect.~\ref{eloss}. The excitation of higher Fock states by multiple interactions leads to 
enhanced nuclear suppression of particle production with large $x_T$ and/or $x_L$
\cite{isi}. This effect can be seen in the $p_T$ dependence of $R_{AA}$ at the RHIC energies $\sqrt{s}=200\GeV$ and $62\GeV$. Also LHC data at $\sqrt{s}=2.76\TeV$ indicate a leveling of the $R_{AA}$ behavior at the maximal measured $p_T$,
and we expect even a fall at $p_T\gsim100\GeV$.

A complementary test of our approach is provided by  data on
azimuthal  anisotropy of produced hadrons, as is described in Sect.~\ref{v2}. In fact, we explain the measured difference between $R_{AA}$ for in- and out-of-plane events and the asymmetry parameter $v_2(p_T)$, with no additional adjustment.

\section{Energy conservation:  How long does hadronization last ?}\label{lp}

\begin{figure}[h]
 \includegraphics[height=3cm]{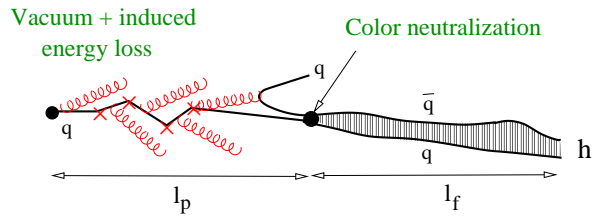}
\caption{ \label{fig:space-time}  (Color online) 
Space-time development of hadronization of a highly virtual quark producing a leading hadron, which carries the main fraction $z_h$ of the initial quark light-cone momentum.
}
 \end{figure}

One should discriminate between the observation of a jet 
initiated by a parton produced in a hard reaction (e.g. high-$p_T$ processes or deep-inelastic scattering (DIS)), and the detection of only a single hadron produced inclusively with a large fractional light-cone momentum in a hard process at high energies.
The latter process corresponds to a very rare jet configuration, 
where  the main fraction $z_h$ of the jet energy $E$ is carried by a single hadron, while all other hadrons in the jet must share the smaller energy $(1-z_h)E$. The deficit of energy imposes certain constraints on the space-time development of such a jet, which is different from an averaged jet. Our definition of the characteristic time scales is illustrated in Fig.~\ref{fig:space-time}.
The quark regenerating its color field, which has been stripped off in a hard reaction, intensively radiates gluons and dissipates energy, either in vacuum or in a medium.
Multiple interactions in the medium induce additional, usually less intensive, radiation.
The loss of energy ceases at the moment,  call production time $t_p$, when the quark picks up an antiquark neutralizing its color. The produced colorless dipole (also called pre-hadron) does not have either the wave function or mass of the hadron, but it takes the formation time $t_f$ to develop both. The formation stage is described within the path-integral method in Sect.~\ref{path}.

\subsection{String model}

The simple example of the consequences of energy conservation is the string (or color flux tube) model  \cite{cnn}. The usual expectation is that the higher is the energy of the produced hadron, the longer it takes to be produced, as follows from 
Lorentz time dilation. On the contrary, in \cite{k-nied} it was found that the production time of a hadron with energy $z_hE$ shrinks down to zero upon reaching the kinematic bound of maximal energy $E$. Indeed, the quark that initiated a jet is losing its energy
with the rate $dE/dt=-\kappa$, where $\kappa\approx 1\GeV/\fm$ is the string tension.
The energy loss comes from the hadronization, which is developed as a series of string breaks
by $\bar qq$ pairs tunneling from the vacuum \cite{cnn}. 

Since the leading quark keeps losing energy, it can produce a hadron with energy $z_hE$ only  within a certain time interval, restricted by energy conservation,
\beq
t_p\leq \frac{E}{\kappa}\,(1-z_h).
\label{100}
\eeq
Such a shrinkage of the production time was explicitly confirmed in Montecarlo models 
\cite{bg,pir}.

The string model is mentioned here only as an example of the constraints imposed on $t_p$ by energy conservation. It should not be taken literally as a hadronization mechanism for
hard reactions, where a highly virtual parton neutralizes its color perturbatively on a short time scale \cite{pert-hadr}.  

At this point we should specify our terminology in order to avoid further confusions.
Indeed, different time scales are debated in the literature, coherence time, formation time, production time.  What we call production time (in Eq.~(\ref{100}) and in what follows),
is the time of color neutralization of the leading quark by an antiquark picked up from the string or generated perturbatively \cite{pert-hadr}. Notice that this is not yet the final hadron, which is characterized by both a specific wave function and mass. What is produced at the time scale $t_p$
is a colorless $\bar qq$ dipole, having certain separation, but not mass, which we call conventionally pre-hadron.
It takes the formation time to develop the wave function,
\beq
t_f=\frac{2E_h}{m_{h^*}^2-m_h^2},
\label{120}
\eeq
where $E_h=z_hE$; $m_h$ and $m_{h^*}$ are the masses of the hadron (we assume it to be the ground state) and the first radial excitation $h^*$.
This time scale does not shrink at $z_h\to1$, but keeps rising. It can be derived in the multi-channel representation for the diffractive scattering as the inverse minimal value of the longitudinal momentum transfer in the off-diagonal diffractive transitions between different states. 

Eq.~(\ref{120}) can also be understood in terms of the uncertainty principle.
Namely, the produced $\bar qq$ dipole has a certain size and can be projected either to $h$, or to $h^*$. According to the uncertainty principle, it takes the proper time $t_f^*=1/(m_{h^*}-m_h)$ to resolve between these two levels.
Applying the Lorentz boosting factor we get (\ref{120}).

Concerning the coherence time scale, this is a more general term, which means that quantum-mechanical interferences are important. Depending on the context and the employed theoretical tools, it might play role on either the production time or the formation time.

\subsection{Radiational energy loss}

As a result of hard interaction characterized with the scale $Q^2$, the parton is produced with part of its color field  stripped off, up to transverse frequencies $k_T\lsim Q$.
Hadronization of such a highly virtual quark cannot be described adequately in terms of the nonperturbative string model. Regeneration of the quark color field is associated with radiation of gluons, which take away a part of the quark energy and contribute to the formation of the jet.  
In fact, at high virtualities $Q^2$ this gluon radiation becomes the dominant source of energy loss
in vacuum.

One should strictly discriminate between vacuum and medium-induced energy loss. The
former includes the lost energy, which goes into gluon radiation and/or into setting up the string field, in other words into jet formation. The latter corresponds to the additional energy loss caused by the multiple interactions of the jet in the medium. The vacuum rate of energy loss usually significantly exceeds the medium-induced one.  
Here we concentrate on the study of the hadronization pattern in vacuum.

The time dependent radiational energy loss can be calculated as follows \cite{knp,within,eloss}
 \beq
\Delta E_{rad}(t) =
E\int\limits_{\lambda^2}^{Q^2}
dk^2\int\limits_0^1 dx\,x\,
\frac{dn_g}{dx\,dk^2}\,
\Theta(t-t^g_c),
\label{140}
 \eeq
 where the coherence time for radiation of a gluon with fractional light-cone momentum $x$ and transverse momentum $k$ reads,
 \beq
t^g_c= \frac{2Ex(1-x)}{k^2+x^2\,m_q^2}.
\label{160} 
\eeq 
The step function in Eq.~(\ref{140}) excludes from the integration those gluons which are still in coherence with the radiation source, and did not materialize on mass shell during the time interval $t$.  The soft cutoff $\lambda$ in Eq.~(\ref{140}) is fixed at $\lambda=0.7\GeV$. This
choice is dictated by data (see in \cite{kst2,spots}), which indicates a
rather large primordial transverse momentum of gluons.

  The spectrum of radiated gluons in Eq.~(\ref{140}) has the form,
 \beq
\frac{dn_g}{dx\,dk^2} =
\frac{2\alpha_s(k^2)}{3\pi\,x}\,
\frac{k^2[1+(1-x)^2]}{[k^2+x^2m_q^2]^2},
\label{180}
 \eeq
 where $\alpha_s(k^2)$ is the running QCD coupling.

The time dependence of radiational energy loss in vacuum exposes a nontrivial behavior \cite{within,eloss}. During the time interval $t<(2E/Q^2)(1-z_h)$ the energy loss rises linearly with time,
\beq
\Delta E_{rad}(t)=t\,\frac{2\alpha_s}{3\pi}\,(Q^2-\Lambda^2)\,.
\label{200}
\eeq
However, at larger $t$ the rate of energy loss starts falling, $\Delta E_{rad}(t)$ is leveling off and at $t>(2E/\Lambda^2)(1-z_h)$ the gluon radiation completely ceases.
Then the quark loses energy only via nonperturbative mechanisms (strings).

Similar to Eq.~(\ref{100}), the production time is restricted by energy conservation to,
\beq
\Delta E(t_p)\approx E(1-z_h).
\label{220}
\eeq

Apparently, the increase  of the energy loss rate in Eq.~(\ref{100}),
$\kappa\Rightarrow\kappa+|dE_{rad}/dt|$, caused by gluon radiation, can only shorten the production time.

\subsection{Peculiar aspects of high-\boldmath$k_T$ jets}

The solution of Eq.~(\ref{220}), the production time of a leading (pre)hadron,
depends on the jet energy and virtuality. In deep-inelastic scattering (DIS) these are two independent variables, and usually $E^2\gg Q^2$.

For a parton produced with high transverse momentum $k_T$ normal to the collision axis in the c.m. frame, its energy
$E\approx k_T$. The hard scale for such a process is also imposed by the transverse momentum,
i.e. $Q^2=k_T^2$. Thus, a high-$k_T$ jet is in an unique kinematic domain of extremely high virtuality, $Q^2=E^2$, which cannot be accessed in DIS.
This fact leads to a specific behavior of the production time for high-$p_T$ hadrons, which is different from what is usually measured in SIDIS.

Indeed, we can trace the dependence of $t_p$ on $E$ and $Q^2$ using the approximate relation analogous to Eq.~(\ref{100}),
\beq
t_p\lesssim \frac{E}{\la |dE/dt|\ra}\,(1-z_h).
\label{221}
\eeq
If one increases the jet energy  keeping the virtuality $Q^2$ fixed, so that $\la |dE/dt|\ra$ does not vary,
the production time rises linearly with $E$ and eventually exceeds the time of jet propagation through the medium. Indeed, the  observed magnitude of nuclear suppression
of leading hadrons in SIDIS decreases with energy and vanishes at jet energies $E\sim100\GeV$ \cite{emc}. However, at medium and low energies \cite{hermes,class} the energy loss scenario \cite{wang-sidis}, which assumes a long production time, fails to describe the nuclear suppression observed in SIDIS at large $z_h$.

And vice versa, if one keeps the energy $E$ fixed, but increases the virtuality $Q^2$,
the mean rate of energy loss in the denominator of (\ref{221}) rises, and the production time shrinks.

Therefore, it is not obvious what will happen to $t_p$ if both the energy and virtuality  rise simultaneously, as it happens for high-$k_T$ jets.  In spite of the rising Lorentz factor, the jet virtuality $Q^2=E^2$ rises as well and  causes a 
dramatic enhancement of radiative energy loss, which may result in a shorter production time. 

Eq.~(\ref{140}) describes the time dependence of the energy radiated by a virtual quark.
The total amount of radiational energy loss is given by the same equation without the $\Theta$-function, 
\beq
\Delta E_{tot}=\frac{8\alpha_s}{3\pi}\,E\,\ln\left(\frac{E}{\lambda}\right),
\label{222}
\eeq
where we assumed $x\ll1$ and fixed $\alpha_s$ for the sake of simplicity.
Let us inverse the problem of the time dependence of energy loss Eq.~(\ref{140}) and ask:
how long does it take for a highly virtual quark or gluon with $Q^2=E^2$ to  radiate a fraction 
\beq
\delta(t)=\frac{\Delta E(t)}{\Delta E_{tot}}
\label{224}
\eeq
of the total radiated energy? Solving  Eq.~(\ref{140}) one gets \cite{trieste},
\beq
t(\delta)\,=\,\left\{
\begin{array}{cl}
\delta\,\frac{4}{E}\,
\ln\left(\frac{E}{\lambda}\right) & 
{\rm if}\ \delta<1/\ln\left(\frac{E^2}{\lambda^2}\right)
\\[4mm]
\frac{2}{\lambda e}
\left(\frac{E}{\lambda}\right)^{2\delta-1} & 
{\rm if}\ \delta>1/\ln\left(\frac{E^2}{\lambda^2}\right)
\end{array}
\right..
\label{230}
\eeq
From the second line of this equation we conclude that a high-$k_T$ parton radiates  half of the total energy loss during a very short time interval, $t(\delta=1/2)=2/(e\lambda)\approx 0.2\fm$. This interval is independent of energy.

\begin{figure}[h]
 \includegraphics[height=6cm]{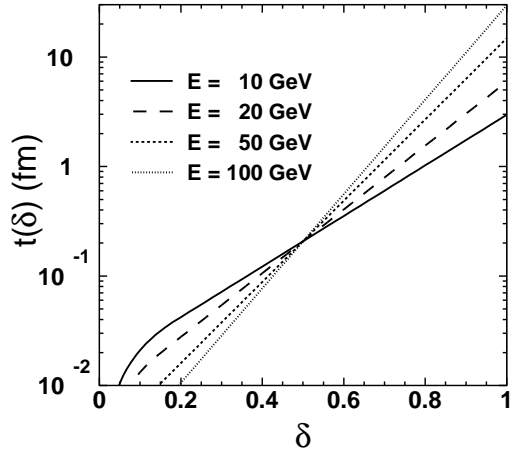}
\caption{ \label{fig:delta-dep} 
The path length taken by a parton (quark or gluon) of energy $E$ and virtuality $Q^2=E^2$ to radiate a
fraction $\delta=\Delta E/E$ of the total vacuum energy
loss. The curves correspond to different
jet energies, $E = 10,\ 20,\ 50$ and $100\GeV$. }
 \end{figure}

The $\delta$-dependence of the time interval $t(\delta)$ is illustrated in Fig.~\ref{fig:delta-dep} for several jet energies.
This confirms that  a high-$k_T$ parton radiates the main fraction of the energy vacuum loss at the early stages of hadronization, on very short time intervals. 
Notice that inclusion of nonperturbative mechanisms of energy loss should lead to even faster dissipation of energy.
Thus, we conclude that the fast degradation of energy of a highly virtual parton
makes impossible the production of leading (pre)hadrons on a long time scale.

\begin{figure}[htb]
\centerline{
  \scalebox{0.37}{\includegraphics{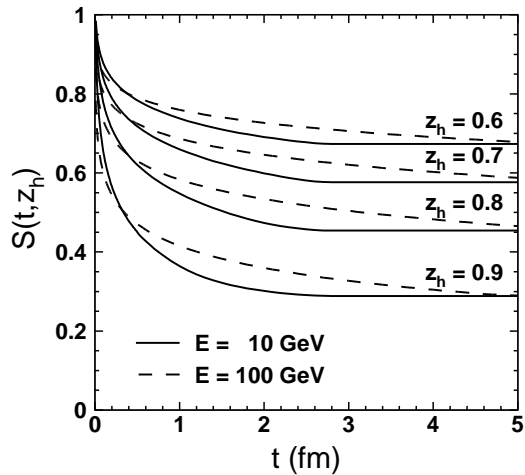}}
 }
\caption{\label{fig:sudakov}
Sudakov suppression factor caused by the ban for radiation of gluons with
fractional energy higher than $1-z_h$, during the time interval $t<t_p=l_p$. Calculations are done for jet energies
$E=Q=k_T=10,\ 100 \GeV$, and several fractional hadron momenta $z_h=0.6-0.9$.
 }
 \end{figure}

So far we explored the time dependence of radiation with no constraints on the radiation process. However, detection of a hadron carrying a large fraction $z_h$ of the initial parton energy essentially affects the energy loss pattern due to energy conservation.
The production of leading hadrons in such jets was studies in \cite{jet-lag}, where it was found to have the form,
 \beq
\frac{\partial D_{\pi/q}(z_h,Q^2)}{\partial t_p}\propto
(1-\tilde z_h)\,S(t_p,z_h)\ .
\label{240}
 \eeq
Here $Q^2=k_T^2$, and the time dependent fractional momentum $\tilde z_h(t)$
reads, 
 \beq
\tilde z_h(t_p) =
\left\la {z_h\over x}\right\ra =
z_h\left(1+\frac{\Delta E(t_p)}{E}\right) +
O\left[z_h(1-z_h)^2\right]\,.
\label{260}
 \eeq
The energy loss $\Delta E(t_p)$ includes both perturbative and nonperturbative (strings) sources of energy dissipation. The former is calculated with Eq.~(\ref{140}), in which an additional kinematical  constraint is introduced: the energy of radiated gluons, $\omega=\alpha E+k^2/4\alpha E$, should not exceed the bound $\omega<(1-z_h)E$.
Such a ban for radiation of part of the gluon spectrum during the time interval $t<t_p$ leads to a suppression known as Sudakov factor $S(t_p,z_h)$, introduced in Eq.~(\ref{240}). It is defined as $S(t_p,z_h)=\exp\left[-\la n_g(t_p,z_h)\ra\right]$, where the mean number of gluons $\la n_g(t_p,z_h)\ra$ which have radiation time shorter than $t_p$, but cannot be radiated due to energy conservation, is calculated with the same gluon spectrum Eq.~(\ref{180}).
Examples for the Sudakov factor at different values of $z_h$ are shown in Fig.~\ref{fig:sudakov}, as functions of the production length $l_p=t_p$ and for jet energies $E=10$ and $100\GeV$.
We see that energy conservation vetoing part of the radiation spectrum results in the Sudakov suppression factor, which substantially reduces the production time Eq.~(\ref{240}).

\begin{figure}[tbh]
 \includegraphics[height=6cm]{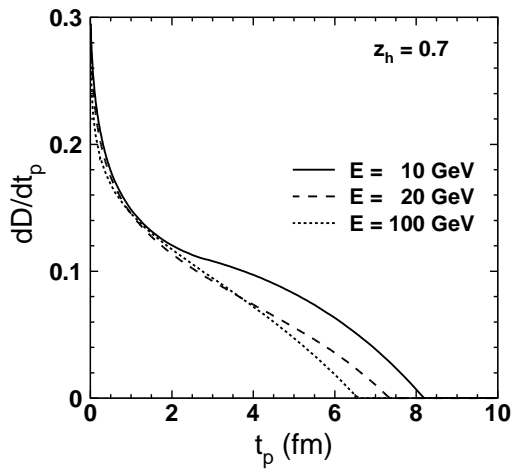}
\caption{ \label{fig:lp-dep} 
The differential fragmentation function Eq.~(\ref{240}) (in arbitrary units)  at $Q^2=E^2$ as function of $l_p$ for quark jets with energies $E=k_T=10,\  20$ and $100\GeV$ (from top to bottom) and $z_h=0.7$. }
 \end{figure}

Now we are in a position to perform numerical calculations for the production time distribution function.
Examples for the differential fragmentation function Eq.~(\ref{240}), at $Q^2=E^2=p_T^2/z_h^2$, are depicted in Fig.~\ref{fig:lp-dep} as function of $l_p$. 
Since the absolute value of the fragmentation function steeply varies with $z_h$, we renormalized it by adjusting to the same value at $l_p=0$, and plot it in arbitrary units.

As a test of  the modeled differential fragmentation function Eq.~(\ref{240}), we integrated it over $t_p$ and compared with data.
Our result reproduces pretty well \cite{jet-lag} the phenomenological function $D_{\pi/q}(z_h,Q^2)$ \cite{bkk} fitted to data, at $z_h\gtrsim0.5$. Correspondingly, we consider our calculated $t_p$-distribution Eq.~(\ref{240}) to be trustable within this interval of $z_h$. 

\begin{figure}[hbt]
 \includegraphics[height=6cm]{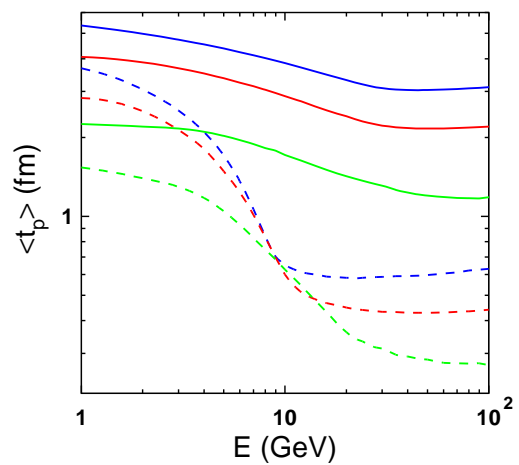}
\caption{ \label{fig:mean-lp}  (Color online) 
The mean production length as function of energy for quark (solid curves) and gluon(dashed curves) jets.
In both cases the curves are calculated at $z_h=0.5,\ 0.7,\ 0.9$ (from top to bottom). }
 \end{figure}

Eventually, using the distribution (\ref{240}) we can calculate the mean production time,
\beq
\la t_p(z_h,E)\ra=\frac{1}{D_{\pi/q}(z_h,E^2)}\int dt_p\,t_p
\frac{\partial D_{\pi/q}(z_h,E^2)}{\partial t_p}.
\label{280}
\eeq
The results are presented in Fig.~\ref{fig:mean-lp}.
Naturally, the production length for leading hadrons in jets initiated by gluons is shorter than for quarks, because the dissipation of energy in gluon jets is more intensive.

Notice that color neutralization, resulting in the production of a pre-hadron, also may be subject to coherence. This means that the amplitudes with different $t_p$ can interfere, so one cannot identify with certainty the production moment $t_p$. One may be even unable to say whether the pre-hadron was created inside or outside the medium. Such a quantum-mechanical uncertainty was explicitly demonstrated in a SIDIS example \cite{avt-tp}, where the interference term in the production cross section was found to be $100\%$ important and negative. However, in the case of a high-$p_T$ jet the starting virtuality is 
so high, and the dissipation of energy so intensive, that every amplitude is constrained to have a short $t_p$. Therefore, in what follows we rely on the  probabilistic description of the space-time development illustrated in Fig.~\ref{fig:space-time}.
  
\subsection{The mean value of \boldmath$z_h$}\label{mean-zh}

As expected, the production time varies with $z_h$, which unfortunately cannot be measured in the process under consideration, but one can evaluate 
its mean value and then rely on it in further calculations. As was stressed above, inclusive production of hadrons with large transverse momentum
enhances the large-$z_h$ part of the fragmentation function $D(z_h,Q^2)$.
This happens due to the steepness of the $k_T$-spectrum of the produced partons,
quarks or gluons, which has to be convoluted with the fragmentation function.
This convolution defines the mean fractional momentum $\la z_h\ra$.

First of all, we should check how well we can describe data in $pp$ collisions. We employ the simple model proposed in \cite{wang}, based on $k_T$-factorization.

\beqn
\frac{d\sigma_{pp}}{dy\,d^2p_T} &=&
K\sum_{i,j,k,l}\,
\int dx_i dx_j d^2k_{iT} d^2k_{jT}
\nonumber\\
&\times&
F_{i/p}(x_i,k_{iT},Q^2)\,
F_{j/p}(x_j,k_{jT},Q^2)
\nonumber\\
&\times&
\frac{d\sigma}{d{\hat t}}(ij\to kl)\,
\frac{1}{\pi\, z_h}\,D_{h/k}(z_h,Q^2).
\label{300}
\eeqn
Here $d\sigma(ij\to kl)/d{\hat t}$ is the cross section of parton scattering; the kinematic variables and their relations can be found in \cite{wang}. Following \cite{wang} we assume a factorized form of the transverse momentum distribution,
\beq
F_{i/p}(x,k_T,Q^2) = 
F_{i/p}(x,Q^2)\,g_p(k_T,Q^2),
\label{320}
\eeq
where
\beq
g_p(k_T,Q^2) =
\frac{1}{\pi \left\la k_T^2(Q^2)\right\ra}\,e^{ - k_T^2/\left\la k_T^2(Q^2)\right\ra}.
\label{330}
\eeq
The scale dependence of $\left\la k_T^2(Q^2)\right\ra$ was parametrized in \cite{wang} as $\la k_T^2\ra_N(Q^2) = 1.2\GeV^2 + 0.2\alpha_s(Q^2)Q^2$, with parameters adjusted to next-to-leading order calculations.

We use the phenomenological 
parton distribution functions (PDF) $F_{i/p}(x,Q^2)$ from MSTW08 leading order (LO) \cite{mstw}. For the fragmentation function $D_{h/k}(z_h,Q^2)$ we rely on the LO parametrization given in \cite{florian}.

\begin{figure}[hbt]
 \includegraphics[height=6cm]{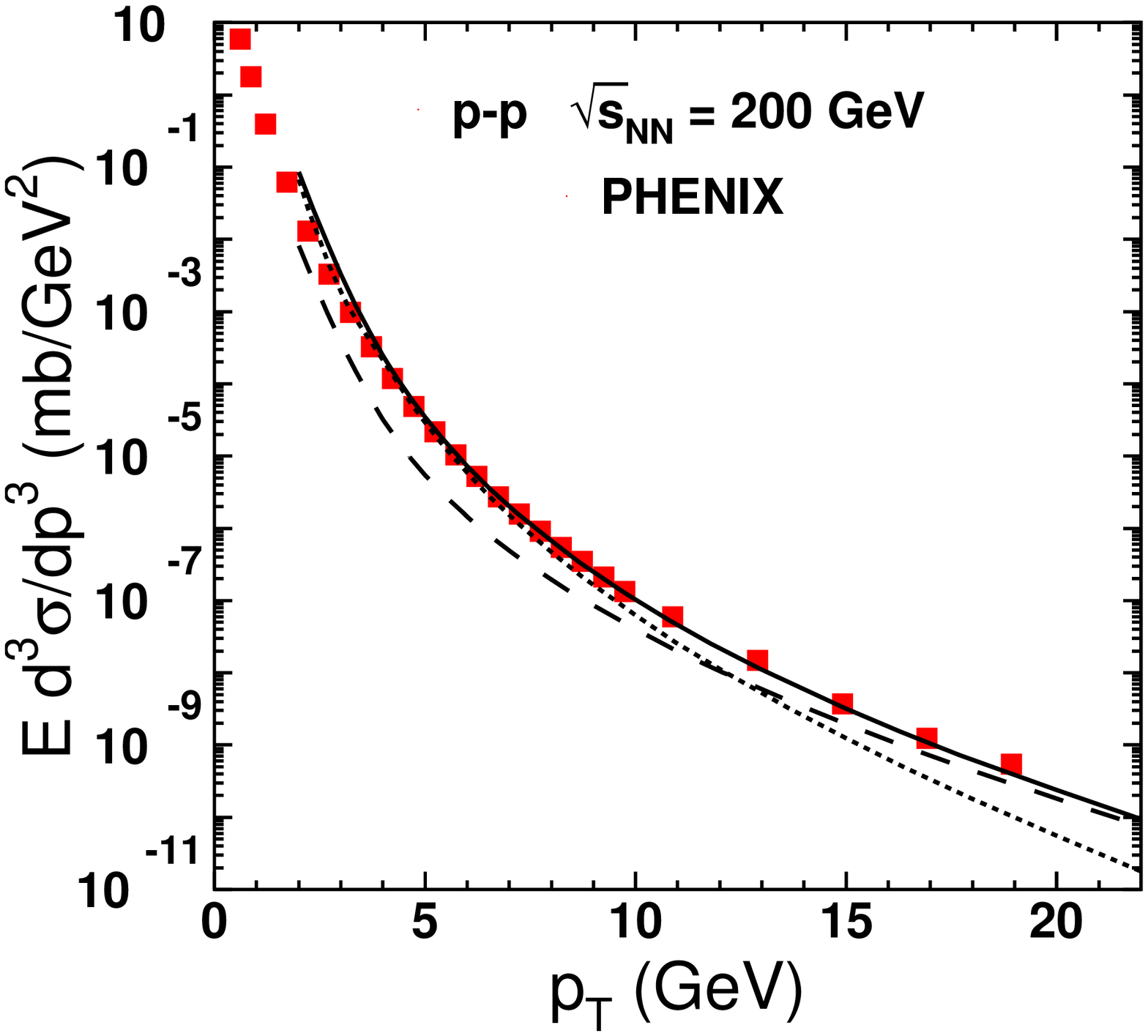}
 \includegraphics[height=6cm]{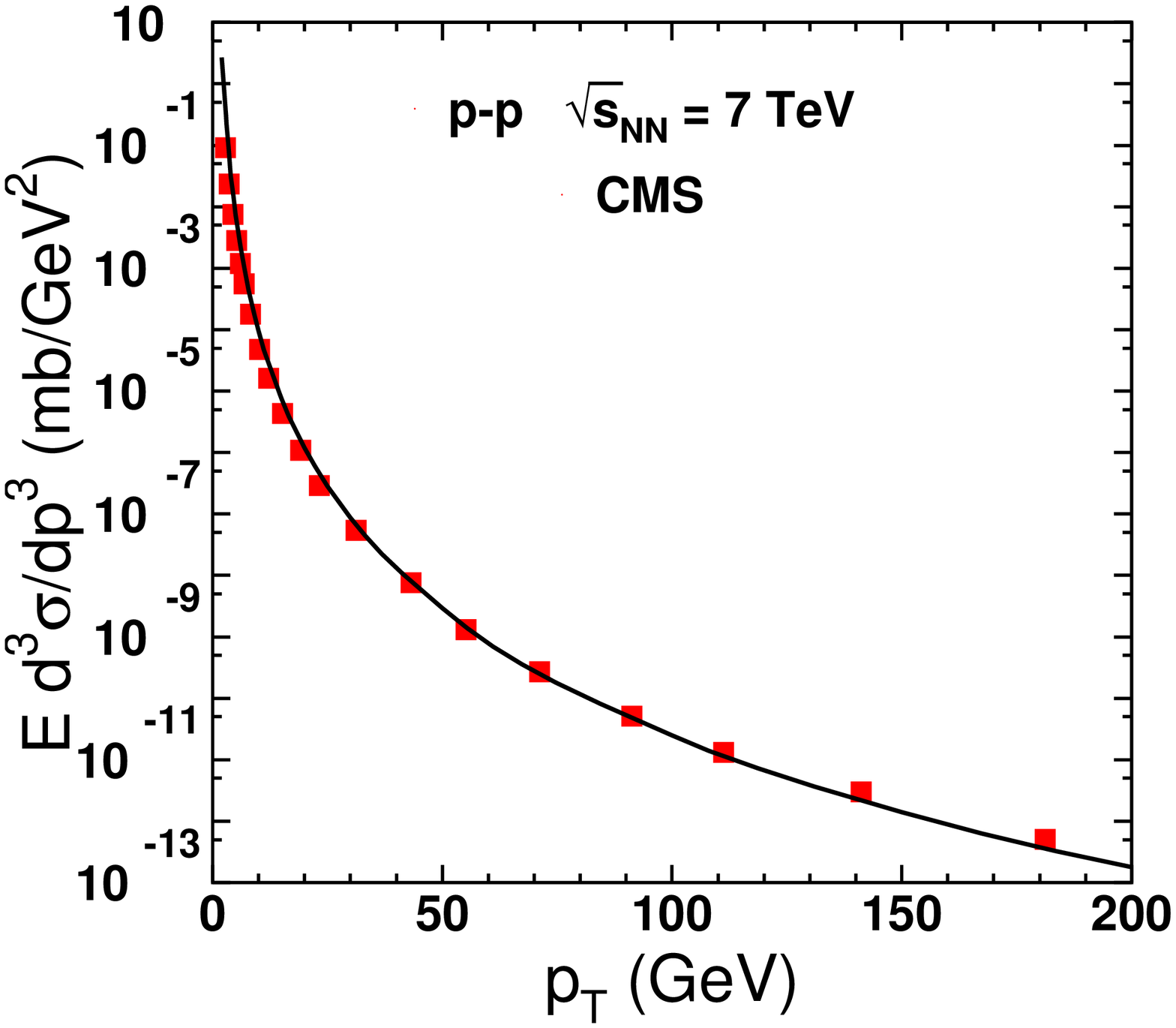}
\caption{ \label{fig:pp-pt-dep}  (Color online) 
$p_T$-dependence of pion production in $pp$ collisions at $\sqrt{s}=200\GeV$ (upper panel) and $7\TeV$ (lower panel). The contributions of quark and gluon
jets are shown in the upper panel by dashed and dotted curves respectively.
Data points are from the PHENIX \cite{phenix-pp} and CMS \cite{cms-pp} experiments.}
 \end{figure}

The results of the differential invariant cross section calculations Eq.~(\ref{300})
are compared with data at $\sqrt{s}=200\GeV$ \cite{phenix-pp} and $7\TeV$ \cite{cms-pp} in Fig.~\ref{fig:pp-pt-dep}.
We see that the employed model reproduces quite well the shape of the measured cross section  up to the maximal available momenta. The absolute normalization (not important for us) is regularized by the K-factor in (\ref{300}), which we found to be $k\approx1-1.5$ depending on the energy.

Notice that while at the LHC energy the cross section is fully dominated by gluon jets, at
$\sqrt{s}=200\GeV$ quarks are important and even dominate towards the upper end of the available range of $p_T$.

\begin{figure}[hbt]
 \includegraphics[height=6cm]{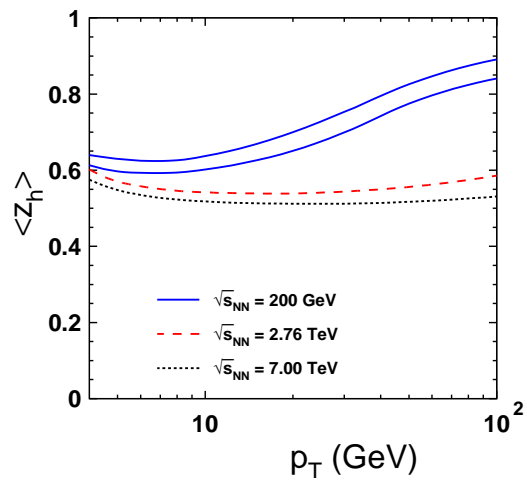}
\caption{ \label{fig:mean-zh}  (Color online) 
The mean fraction $\la z_h\ra$  of the jet energy carried by a hadron detected with transverse momentum $p_T$. The calculations are performed as described in the text, for collision energies $\sqrt{s}=200,\ 2760$ and $7000\GeV$.}
 \end{figure}

Eventually, we can average the fractional hadron momentum $z_h$ 
weighted with the convolution Eq.~(\ref{300}). The results are depicted in Fig.~\ref{fig:mean-zh}, separately for quark and gluon jets (upper and bottom solid curves) and at different energies, $\sqrt{s}=200,\ 2760$ and $7000\GeV$.
We see that the lower is the collision energy, the larger is $\la z_h\ra$, especially at large $p_T$, because the parton $k_T$ distribution gets steeper. In the energy range of the LHC the magnitude of $\la z_h\ra$ practically saturates as function of $\sqrt{s}$ and $p_T$.

Such a large value of the fractional jet energy carried by a single  inclusively detected high-$p_T$ hadron (without observation of the whole jet), makes its space-time development very different from an usual high-$p_T$ jet, when the whole jet is reconstructed. In the latter case,
if no special selection is made, the mean fractional momenta of hadrons are very small.
Correspondingly, energy conservation does not impose any severe constraints on the hadronization time scale, which rises with $p_T$ and may be long.

\section{Attenuation of leading hadrons in a dense medium}\label{dipole}

In previous sections we found that in rare events, in which the produced hadron carries the main fraction $z_h$ of the initial parton light-cone momentum, the intensive gluon radiation and energy dissipation in vacuum by a highly virtual
parton produced with high $k_T$ does not leave much time for the hadronization process. Particularly, such rare events are selected by detecting
a hadron with large $p_T$. Fig.~\ref{fig:mean-zh} shows that the detected hadron carries on average more than half of the jet momentum.
In order to respect energy conservation, the intensive dissipation of energy by the parton should be stopped promptly by color neutralization, i.e. production of a colorless pre-hadron, otherwise the leading parton, which lost too much energy, will be unable to produce a hadron with large $z_h$.  Other possibilities for reduction of the vacuum energy loss by non-radiation of gluons are strongly Sudakov suppressed.
As a result, the time scale for production of a colorless dipole is rather short and does not rise with $p_T$, as is depicted in Fig.~\ref{fig:mean-lp}.

If this process occurs not in vacuum, but in a dense medium, multiple interactions of the parton generate more energy loss, which makes the production time even shorter.
The further interactions of the dipole in the medium are mostly inelastic color exchange collisions.
Indeed, the cross section of inelastic interactions is proportional to the dipole separation squared, $r_T^2$, while the elastic scattering cross section  
is $\propto r_T^4$, which  is negligibly small. So, any inelastic interaction of the colorless dipole with color exchange will resume the gluon radiation and dissipation of energy. Of course color neutralization  may happen again via creation of a new dipole, but such a reincarnation of the pre-hadron will result in a substantial reduction of its momentum.

Thus, we should evaluate the survival probability $W$, i.e. the chance for a dipole to escape from the dense medium having no inelastic interaction on the way out. Apparently, this is subject to the effect of color transparency \cite{zkl}, i.e. the rate of attenuation of small-size dipoles
vanishes quadratically with the dipole transverse separation $r$, 
\beq
\left.\frac {dW}{dl}\right|_{r_T\to0}=-\epsilon(l)\,r^2,
\label{302}
\eeq
where $\epsilon(l)$ characterizes the medium, and varies with coordinates and time.

To avoid terminological confusions, notice that in high-$p_T$ hadron production the momentum 
$\vec p_T$ is meant to be transverse to the collision axis.  However, the ``transverse"
dipole separation is meant to be transverse relative to the vector $\vec p_T$.

It was found in \cite{jkt} that  a parton propagating through a nuclear medium  experiences  broadening, whose magnitude is controlled by the small-$r$ behavior of the dipole-nucleon cross section.
Thus, the broadening and attenuation rates in a medium turn out to be related as,
\beq
\epsilon(l)={1\over2}\,\hat q(l)
\label{322}
\eeq
Here the broadening rate $\hat q(l)=\partial\Delta q^2/\partial l$ is usually called transport coefficient and used as a characteristics of the medium \cite{bdmps}. Notice that $\vec q$ is the transverse momentum of the parton relative to its initial direction.
The transport coefficient is proportional to the medium density, which is function of impact parameter and time. In what follows we rely on the popular, although poorly justified model for $\hat q$, which is assumed to be proportional to the number of participants and gets
diluted with time as  $\rho(t)=1/t$, due to the
longitudinal expansion of the produced medium. 
Correspondingly, the transport coefficient depends on impact parameter and time (path length $l=t$) as \cite{frankfurt},
\beq 
\hat q(l,\vec b,\vec\tau)=\frac{\hat
q_0\,l_0}{l}\, \frac{n_{part}(\vec b,\vec\tau)}{n_{part}(0,0)}
\,\Theta(l-l_0),
\label{340} 
\eeq
where $\vec b$ is the impact parameter of nuclear collision, $\vec\tau$ is the impact parameter of the hard parton-parton collision relative to the center of one of the nuclei,,$n_{part}(\vec b,\vec\tau)$ is the number of participants, and
$\hat q_0$ is the rate of broadening of a quark propagating in the 
maximal medium density produced at impact parameter $\tau=0$ in
central collisions ($b=0$) at the time $t=t_0=l_0$ after
the collision. The corresponding transport coefficient for gluons should be $9/4$ bigger.
The equilibration time $t_0$ is model dependent. Our results are not very sensitive to it, 
and we fix it at $t_0=l_0=0.5\fm$.

As far as the process of high-$p_T$ hadron production is considered as a probe for the medium properties, we  treat the transport coefficient 
$\hat q(l,\vec b,\vec\tau)$ as an adjustable quantity. Once the shape of this function is fixed by the model Eq.~(\ref{340}), the only fitted parameter 
is the maximal value of the transport coefficient $\hat q_0$, which is $A$-dependent.

\subsection{Evolution and attenuation of a dipole:\\ heuristic description}
\label{simple}

Here we employ a simplified description of the time evolution, in terms of the mean dipole transverse separation. The dipole produced with a very small initial size
$r\sim 1/k_T$ starts expanding with a speed given by the uncertainty relation
$dr/dt\propto1/r$ \cite{psi,psi-bnl,my-alice}.
Correspondingly, the $l$-dependence of $r$ is described by the following linear differential equation
 \footnote{Factor 2 was missed in \cite{my-alice}},
\beq
\frac{dr}{dl}=\frac{1}{r(l)E_h\alpha(1-\alpha)},
\label{360}
\eeq
where $\alpha$ and $1-\alpha$ are the fractions of the dipole light-cone momentum carried by the quark and antiquark, and $E_h=p_T$ is the energy of the dipole.

The solution of Eq.~(\ref{360}) reads,
\beq
r^2(l) = \frac{2l}{\alpha(1-\alpha) p_T}+r_0^2,
\label{380}
\eeq
where $r_0$ is the initial dipole size. If it is small, $r_0\sim1/k_T$, its magnitude is quickly ``forgotten". Indeed, the first term in (\ref{380}) starts dominating at $l\gg1/p_T$, and the value of the initial size $r_0$ is unimportant.

The mean value of $r^2(l)$ Eq.~(\ref{380}) can be used in Eq.~(\ref{302}) to evaluate the attenuation of a dipole,  initially small and evolving its size in a medium characterized with a transport coefficient $\hat q$. 
\beq
R_{AB}(\vec{b},\vec\tau,p_T) =
\int\limits_0^{2\pi}
\frac{d\phi}{2\pi} \exp\Biggl[-\frac{4}{p_T}\!
\int\limits_L^\infty \!\!dl \,l\,\hat q(l,\vec{b},\vec\tau+\vec l)
\Biggr].
\label{400}
\eeq
This is the medium attenuation factor for a dipole produced in a hard internal $NN$ collision at impact parameter $\vec\tau$,  in a collision of nuclei $A$ and $B$
with impact parameter $\vec b$. The produced hadron is detected at azimuthal angle $\phi$ relative to $\vec b$, i.e. $\vec l\cdot\vec b=lb\cos\phi$.
The bottom limit of $l$-integration is $L=\max\{l_p,l_0\}$. The dependence of the transport coefficient on coordinates is given by Eq.~(\ref{340}).
In (\ref{400}) we fixed $\alpha=1/2$, because the dipole distribution amplitude over $\vec r$ and $\alpha$ is projected to the hadron (pion) wave function, which has a maximum at $\alpha=1/2$ \cite{radyushkin}. Moreover, we rely on Berger's approximation \cite{berger}, which fixes  $\alpha$ at this value.

This simplified approach was employed in \cite{my-alice} and described quite well the
first data from the ALICE experiment \cite{alice-data} for central collisions.
Important observations made in \cite{my-alice} are: (i) the value of the parameter $\hat q_0$ needed to explain the observed suppression is an order of magnitude smaller than the results of the data analysis \cite{phenix-theor}, based on the energy-loss scenario. At the same time, it agrees well with the perturbative evaluation of $\hat q_0$ \cite{bdmps}, and with the analysis \cite{psi,psi-bnl} of data on suppression of $J/\Psi$ produced in central gold-gold collisions at $\sqrt{s}=200\GeV$.
(ii) The observed rising $p_T$-dependence of $R_{AA}(p_T)$ \cite{alice-data} is naturally explained by the effect of color transparency Eq.~(\ref{302}). Namely, the higher is the dipole energy $E_h=p_T$ (in the medium rest frame), the more is the Lorentz dilation of the dipole size expansion, and therefore the less is the absorption. This is why there is a factor $1/p_T$ in the exponent of Eq.~(\ref{400}).

As was commented above, this simplified heuristic model allows to easily understand the main features of the underlying dynamics. It also allows to speed up the calculations.  However, it misses some details which may affect the result. In particular, Eq.~(\ref{360}) describes the expansion of the dipole in vacuum. However, in a medium, color filtering effects modify the path-length dependence of the mean dipole separation. Namely,
dipoles of large size are strongly absorbed, while small dipoles attenuate less.
Correspondingly, the mean separation in a dipole propagating in a medium should be smaller than that given by the differential equation (\ref{360}) for the dipole expansion in vacuum. Introducing an absorptive term we arrive at a modified evolution equation,
\beq
\frac{dr^2}{dl}=\frac{2}{E_h\alpha(1-\alpha)}-r^4(l)\,\epsilon(l),
\label{410}
\eeq
where $\epsilon(l)$ is related by (\ref{322}) to the transport coefficient.
Apparently, such a modification results in a reduction of the mean dipole size, making the medium more transparent. Correspondingly, we should expect that the analysis of ALICE data, performed in \cite{my-alice}, should have underestimated the medium density, i.e.
the parameter $\hat q_0$. We will return to this problem in Sect.~\ref{RAA}.

Unfortunately, Eq.~(\ref{410}) has an analytic solution only if $\epsilon(l)$ is constant,
which is not the case here. Then one should solve the equation numerically, so the simplicity of such a heuristic model is lost. In these circumstances it is worth
switching to the rigorous
quantum-mechanical description of the evolution and attenuation of a dipole in an absorptive medium, and employ the path-integral approach \cite{kz91}.

\subsection{Path-integral technique}\label{path}

Evolution and attenuation of a $\bar qq$ dipole propagating through an absorptive medium, starting 
from the transverse $\bar qq$ separation $\vec r_1$ at a point with longitudinal coordinate $l_1$ and  evolving its size up to $\vec r_2$ at the point $l_2$, is given by a sum over all possible trajectories of $q$ and $\bar q$.
The resulting survival probability amplitude has the form of a light-cone Green function $G_{\bar qq}(l_1,\vec r_1;l_2,\vec r_2)$, which satisfies the two-dimensional Schr\"odinger equation
\cite{kst1,kst2,krt,zakharov},
\beqn
\left[i\frac{d}{dl_2}\!\!\right.&-&\left.\!\!
\frac{m_q^2 - \Delta_{r_2}}{2\,p_T\,\alpha\,(1-\alpha)}
-V_{\bar qq}(l_2,\vec r_2)\right]
G_{\bar qq}(l_1,\vec r_1;l_2,\vec r_2)
\nonumber\\
&=& i\delta(l_2-l_1)\,\delta(\vec r_2-\vec r_1),
\label{420}
\eeqn
and the boundary conditions,
\beqn
G_{\bar qq}(l_1,\vec r_1;l_2,\vec r_2)\Bigr|_{l_1=l_2} &=&
\delta(\vec r_2-\vec r_1);\nonumber\\
G_{\bar qq}(l_1,\vec r_1;l_2,\vec r_2)\Bigr|_{l_1>l_2} &=& 0
\label{440}
\eeqn
The  second term in square brackets in (\ref{420}) plays role of the kinetic energy in Schr\"odinger equation, while the imaginary part of the light-cone potential $V_{\bar qq}(l_2,\vec r_2)$ in (\ref{420}) is responsible for absorption in the medium. According to Eqs.~(\ref{302})-(\ref{322}),
\beq
\Im V_{\bar qq}(l,\vec r) = -{1\over4}\,\hat q(l)\,r^2.
\label{460}
\eeq

The real part of the potential describes the nonperturbative interaction between $q$ and $\bar q$ in the dipole \cite{kst2,VM}.
It is questionable, however, whether such a binding potential should be considered within a hot, maybe de-confined medium. Therefore we will treat the $\bar qq$ as free noninteracting partons, like in the previous Sect.~\ref{simple}. The real potential
should not affect much the dipole evolution on the initial perturbative stage of development.

In the case of a constant medium density $\hat q(l)\!=\!\hat q\!=\!const$,  Eq.~(\ref{420})
allows an analytic solution \cite{kz91},
\begin{widetext}
\beq
G_{\bar qq}(l_1,\vec r_1;l_2,\vec r_2) =
\frac{\gamma}{2\pi i
\sin(\omega \Delta l)}\,
\exp\left\{\frac{i\,\gamma}{2\sin(\omega\Delta l)}
\Bigl[(r_1^2+r_2^2)\cos(\omega \Delta l) -
2\vec r_1\cdot\vec r_2\Bigr]\right\}\,
{\exp}\left[-
\frac{im_q^{2}\Delta l}
{2p_T\alpha(1 - \alpha)}\right]
\label{480}
\eeq

\end{widetext}

where $\Delta l = l_2 - l_1$ and
 \beqn
\omega^2 &=& 
 - {i\over2}\,\frac{\hat q}
{p_T\alpha (1 - \alpha)};
\nonumber\\
\gamma^2 &=& 
- {i\over2}\,p_T\alpha (1 - \alpha)\hat q.
\label{500}
 \eeqn

It was demonstrated in \cite{kz91,preasymptotics} that in the general case of arbitrarily varying medium
density the Green function still retains the oscillatory form,
\beqn
&&
\!\!\!\!\!G_{\bar qq}(l_1,\vec r_1;l_2,\vec r_2) =
\Omega(l_1,l_2)
{\exp}\left[-
\frac{im_q^{2}\Delta l}
{2p_T\alpha(1 - \alpha)}\right]
\label{520}\\ &\times&
\!\!\exp\left\{\Upsilon_1(l_1,l_2)\,r_1^2+
\Upsilon_2(l_1,l_2)\,r_1^2+
\Upsilon_3(l_1,l_2)\,\vec r_1\cdot\vec r_2\right\}.
\nonumber
\eeqn
All information about the dipole absorption rate varying along its trajectory, $\hat q(l)$,
is contained in the coefficients $\Omega$ and $\Upsilon_i$. They are calculated numerically by slicing the medium into layers, which are sufficiently thin to keep $\hat q$, given by (\ref{340}), constant within each of them. Then one can calculate the coefficients $\Omega(l_1,l_2)$ and $\Upsilon_i(l_1,l_2)$, employing the recurrence relations between subsequent layers, derived in \cite{kz91, preasymptotics}.

As in the previous section, we rely on the Berger model \cite{berger}, which assumes equal sharing of the pion light-cone momentum between $q$ and $\bar q$, i.e. $\alpha=1/2$. Since the Green function is projected into the pion light-cone wave function, we have to keep $\alpha=1/2$ for the dipole as well.

Now we are in a position to write a rigorous quantum-mechanical extension of the simplified model (\ref{400}), for the suppression factor $R_{AB}(\vec b)$ of hadrons produced with high $p_T$ in a hard process in a
collision of nuclei $A$ and $B$ with relative impact parameter $b$,
\begin{widetext}
\beqn
R_{AB}(\vec{b},p_T) &=& 
\frac{\int d^2\tau \,T_A(\tau)T_B(\vec b-\vec\tau)
\int\limits_0^{2\pi}\frac{d\phi}{2\pi} 
\left|
\int\limits_0^1d\alpha\int d^2r_1 d^2r_2\,
\Psi_h^\dagger(\vec r_2,\alpha)
G_{\bar qq}(l_1,\vec r_1;l_2,\vec r_2)
\Psi_{in}(\vec r_1,\alpha)\right|^2
}{T_{AB}(b)
\left|\int\limits_0^1d\alpha\int d^2r\,
\Psi_h^\dagger(\vec r_2,\alpha)
\Psi_{in}(\vec r_1,\alpha)\right|^2
}
\nonumber\\ &=&
\frac{1}{T_{AB}(b)
\left|\Psi_h(0)\right|^2}
\int d^2\tau \,T_A(\tau)T_B(\vec b-\vec\tau)
\int\limits_0^{2\pi}d\phi 
\left|\int\limits_0^\infty dr\,r\,\Psi_h(r)\, 
G_{\bar qq}(0,0;l_{max}, r)\right|^2,
\label{540}
\eeqn
\end{widetext}
where $T_{AB}=\int d^2\tau\,T_A(b)T_B(\vec b-\vec\tau)$;
$\phi$ is the azimuthal angle of the dipole trajectory in impact parameter plane, relative to the impact vector $\vec b$ of the collision. $l_{max}$ is any distance, which should be much longer that the extent of the medium. Its particular length is unimportant, because the Green function in vacuum is just a phase. Notice that all information about the dipole trajectory, including the $\phi$-, $\vec\tau$- and $\vec b$-dependences, is contained in the Green function.
In (\ref{540}) we fixed $r_0=0$, because according to the solution Eq.~(\ref{380}), it does make a difference whether $r_0$ is as small as $1/k_T$ or zero.
It is also worth noting that we start the evolution of the Green function $G_{\bar qq}(l_1,\vec r_1;l_2,\vec r_2)$ at $l_1=0$, i.e. at the point of the hard collision, but assume that the absorptive imaginary
part of the potential Eq.~(\ref{420}) is zero at $l<l_0$ (compare with (\ref{340})). 
Although the dipole does not exist at $l<l_p$, the parton virtuality is steeply falling, 
governed by the same equation (\ref{360}), and the dipole is produced at $l=l_p$ 
with an enlarged separation, the same as if it had started evolution at $t=0$.
We remind that the mean production length $\la l_p\ra$ is different for quark and gluon jets as is demonstrated in Fig.~\ref{fig:mean-lp}. Therefore the numerator in (\ref{540})
is calculated separately for quark and gluon jets and then summed with the weights given by Eq.~(\ref{300}).

During the short path from $l=l_0$ to $l=l_p$ (if $l_p>l_0$) the parton experiences
multiple interactions, which induce extra radiation of gluons and additional loss of energy
\cite{bdmps},
\beq
\Delta E=\frac{3\alpha_s}{4}\,
\Theta(l_p-l_0)
\int\limits_{l_0}^{l_p} dl
\int\limits_{l_0}^l dl'\,\hat q(l').
\label{550}
\eeq

Although this is a small correction, we included it in the calculations by making a proper shift of the variable $z_h$ in the fragmentation function.

\section{Comparison with data}\label{data}

Numerous results of new precise measurements at RHIC and LHC have been released recently. They allow to perform stringent tests of the contemporary models of in-medium hadronization.

\subsection{Quenching of high-\boldmath$p_T$ hadrons}\label{RAA}

The comparison of $R_{AA}(b,p_T)$ calculated within the simple model described in Sect.~\ref{simple}, with data from the ALICE experiment \cite{alice-data} at $\sqrt{s}=2.76\TeV$, was performed in \cite{my-alice}. While the absolute value of $R_{AA}$
is adjustable, its rising $p_T$-dependence originates from the reduction of the mean dipole size with $p_T$, in accordance with Eq.~(\ref{380}), and due to Lorentz dilation of the dipole size expansion. As a result, the medium becomes more transparent for more energetic dipoles in accordance with the effect of color transparency (CT), which makes the medium more transparent for smaller dipoles. An analogous rising energy dependence of medium transparency was predicted and observed for virtual photoproduction of vector mesons on nuclei \cite{CT}.
So it was concluded in \cite{my-alice} that CT is the source of the rising $p_T$ dependence of $R_{AA}$ observed in the ALICE experiment \cite{alice-data}.

\begin{figure}[htb]
\vspace*{5mm}
\centerline{
  \scalebox{0.4}{\includegraphics{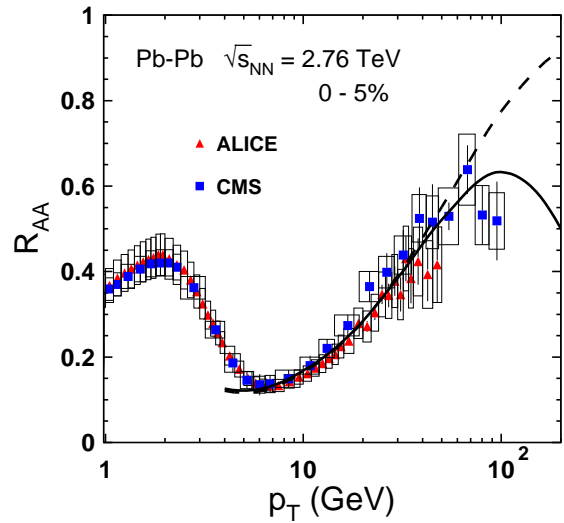}}}
\caption{\label{fig:lhc-0-5-gf} (Color online) The suppression factor $R_{AA}$ for central ($0-5\%$) lead-lead collisions at $\sqrt{s}=2.76\TeV$. The dashed line is calculated within the path-integral approach, Eq.~(\ref{540}) with the space- and time dependent transport coefficient Eq.~(\ref{340}), where the adjusted parameter  $\hat q_0=2\GeV^2\!/\!\fm$.
The solid curve also includes the effects of initial state interactions in nuclear collisions \cite{isi,kn-review}, 
as is described in Sect.~\ref{eloss}. Data for $R_{AA}$ are from the ALICE \cite{alice-new} and CMS \cite{cms-new1,cms-new2} experiments.}
 \end{figure}

Encouraged by the success of the simple model,
we perform here full calculations employing the path integral method, Eq.~(\ref{540}).
The results for central ($0-5\%$) lead-lead collisions at $\sqrt{s}=2.76\TeV$ are shown by the dashed curve in Fig.~\ref{fig:lhc-0-5-gf}, compared with new data from the ALICE \cite{alice-new} and CMS \cite{cms-new1,cms-new2} experiments, extended to higher values of $p_T$ than those in \cite{alice-data}. 
The only free parameter, the maximal transport coefficient defined in Eq.~(\ref{340}),
was adjusted to the data and fixed at $\hat q_0=2\GeV^2\!/\!\fm$ for all further calculations for lead-lead collisions at this energy.

Notice that this value of $\hat q_0$ is about twice larger than what was found in \cite{my-alice} within the simple model described above in Sect.~\ref{simple}. As we commented, this should be expected, since Eq.~({\ref{360}) is lacking the color filtering effect added in
(\ref{410}) or inherited from the Green function equation (\ref{420}).

While our calculations describe well the data at high $p_T\gsim 6\GeV$, the region of smaller $p_T$ is apparently dominated by a thermal mechanisms of hadron production, as is confirmed by the large elliptic flow observed in this region (see below).
This is why we do not try to explain data in this region by the perturbative dynamics.

The variation of the suppression factor $R_{AA}(p_T,b)$ with impact parameter of collision
was also calculated with the Eq.~(\ref{540}). The results plotted by dashed curves are compared with
data taken at different centralities of collision by the ALICE experiment \cite{alice-new}
in Fig.~\ref{fig:alice-b-gf}, and by the CMS experiment \cite{cms-new1,cms-new2} in Fig.~\ref{fig:cms-b-gf}. 
In all cases we observe good agreement.

 \begin{figure}[htb]
\vspace*{5mm}
\centerline{
  \scalebox{0.45}{\includegraphics{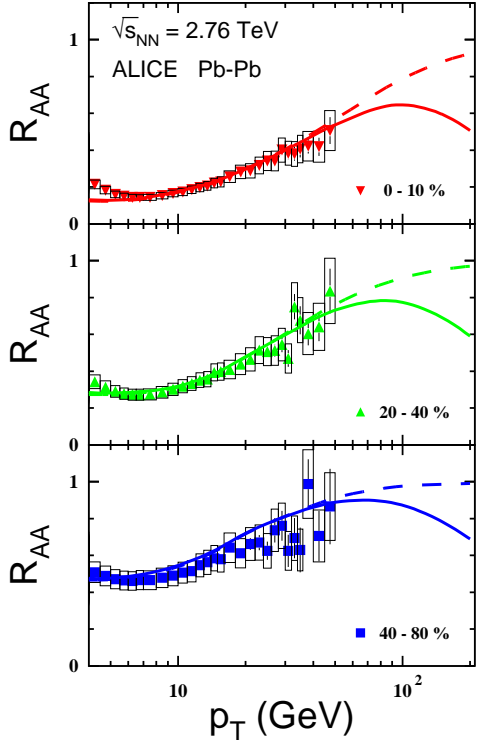}}}
\caption{\label{fig:alice-b-gf} (Color online) Centrality dependence of the suppression factor $R_{AA}(p_T,b)$ measured in the ALICE experiment \cite{alice-new}. The intervals of centrality are indicated in the plot. The meaning of the curves is the same as in Fig.~\ref{fig:lhc-0-5-gf}. }
 \end{figure}

 \begin{figure}[htb]
\vspace*{5mm}
\centerline{
  \scalebox{0.45}{\includegraphics{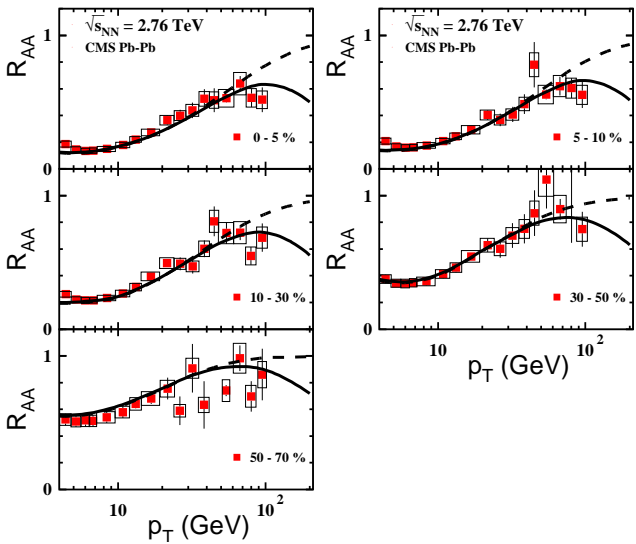}}}
\caption{\label{fig:cms-b-gf} (Color online) The same as in Fig.~\ref{fig:alice-b-gf},
with data from the CMS experiment \cite{cms-new1,cms-new2}. }
 \end{figure}

\subsection{Towards large \boldmath$x_T$: new constraints from\\ energy conservation}
\label{eloss}

It was stressed in \cite{isi,kn-review} that energy conservation may become an issue
upon approaching the kinematic bound of either large Feynman $x_L\eqq x_F\eq2 p_L/\sqrt{s}$, or transverse fractional momentum $x_T \eq 2p_T/\sqrt{s}$. Apparently, in such a kinematic domain any initial state interactions (ISI) leading to energy dissipation, should result in a suppressed production rate of particles with large $x_{L,T}$. Indeed, as it was stressed in \cite{isi},   {\it every reaction} experimentally studied so far, with any particle (hadrons, Drell-Yan dileptons, charmonium, etc.) produced with large $x_L$, exhibits nuclear suppression. Observation of such a suppression in high $x_T$ processes is more difficult, because the cross sections steeply fall with $p_T$, and available statistics
may not be sufficient (see however below Fig.~\ref{fig:cronin-c}). As in \cite{kn-review}, we apply to high-$p_T$ production exactly the same model developed in \cite{isi,isi-jan} for large $x_L$, and with the same parameters.

Since initial state multiple collisions suppress the production rate of leading particles, we assume that every collision brings in a 
suppression factor $U(\xi)$ \cite{isi}, where
$\xi=\sqrt{x_L^2+x_T^2}$
This factor $U(\xi)$ should cause a strong (for heavy nuclei) suppression at $\xi\to1$, although at the same time some 
enhancement at small $\xi$ due to the feed down from higher $\xi$. This is because energy 
conservation does not lead to disappearance (absorption) of particles, but only to their 
re-distribution in $\xi$.
 
 Since at $\xi\to1$ the kinematics of an inelastic collision corresponds to no particle production  within 
the rapidity interval $\Delta y\sim-\ln(1-\xi)$, the suppression factor $U(\xi)$ can also be treated 
as survival probability of a large rapidity gap, which is Sudakov suppressed. The mean number $\la n_g(\Delta y)\ra$ of gluons radiated in the rapidity
interval $\Delta y$ is related to the height of the plateau in the gluon
spectrum, $\la n_g(\Delta y)\ra=\Delta y\,dn_g/dy$. Then, the Sudakov
factor reads,
 \beq
U(\xi) = (1-\xi)^{dn_g/dy}\ .
\label{560}
 \eeq
 The height of the gluon plateau was estimated in
\cite{gb} as,
 \beq
\frac{dn_g}{dy} = \frac{3\alpha_s}{\pi}\,
\ln\left(\frac{m_\rho^2}{\Lambda_{QCD}^2}\right)\ .
\label{580}
 \eeq
 The value of $\alpha_s$ was fitted in \cite{gb} at  $\alpha_s=0.45$, using data on pion multiplicity in $e^+e^-$ annihilation. 
This leads with a good accuracy to $dn_g/dy\approx1$, i.e. the Sudakov factor,
\beq
U(\xi)=1-\xi. 
\label{590}
\eeq

Although QCD factorization is expected to be broken by ISI in $pA$ collisions at large $\xi$, we will rely on the effective factorization formula, Eq.~(\ref{300}), where we replace the proton
parton distribution function by a nuclear modified one $F_{i/p}(x_i,Q^2)\Rightarrow F^{(A)}_{i/p}(x_i,Q^2,b)$. In the case of nuclear collisions we do this modification for the bound nucleons in both nuclei.
Relying on the suppression factor Eq.~(\ref{590}), and applying the AGK cutting rules \cite{agk} with the Glauber weight factors, one arrives at the ISI modified parton distribution function of the proton in a $pA$ collision at impact parameter $b$,
 \beqn
\hspace*{-0.40cm}
F^{(A)}_{i/p}(x_i,Q^2,b)&=&C\,F_{i/p}(x_i,Q^2)\,
\nonumber\\ 
&\times&
\frac{
\left[e^{-\xi\sigma_{eff}T_A(b)}-
e^{-\sigma_{eff}T_A(b)}\right]}
{(1-\xi)\,\left[1-
e^{-\sigma_{eff}T_A(b)}\right]}.
\label{600}
 \eeqn
Here $\sigma_{eff}$ is the effective hadronic cross section controlling multiple interactions. 
It is reduced by Gribov inelastic shadowing, which makes the nuclear medium much more transparent. The effective cross section was evaluated  in \cite{isi,lrg} at about $\sigma_{eff}\approx 20\mb$.
 The normalization factor $C$ in Eq.~(\ref{600}) is fixed by the Gottfried
sum rule, because the number of valence quarks, dominating at large $\xi$, should be unchanged.
  
With the parton distribution functions Eq.~(\ref{600}) modified by ISI
one achieves a good parameter-free description of available data at large $x_L$ \cite{isi,kn-review}. Moreover, these corrections may be important at large $p_T$, in particular in the RHIC energy range, where $x_T$ reaches values of $0.2-0.3$. 
Notice that the real values of $x_T$, essential for energy conservation, are significantly higher,
$\tilde x_T=x_T/z_h$, so it reaches values of $0.3-0.5$ at RHIC, and about $0.4$
at LHC ($p_T=200\GeV$).

Comparison with PHENIX data
\cite{cronin-phenix} for neutral pion production in central $d$-$Au$ collisions is presented in Fig.~\ref{fig:cronin-c}.
\begin{figure}[thb]
\begin{center}
\includegraphics[width=7.0cm]{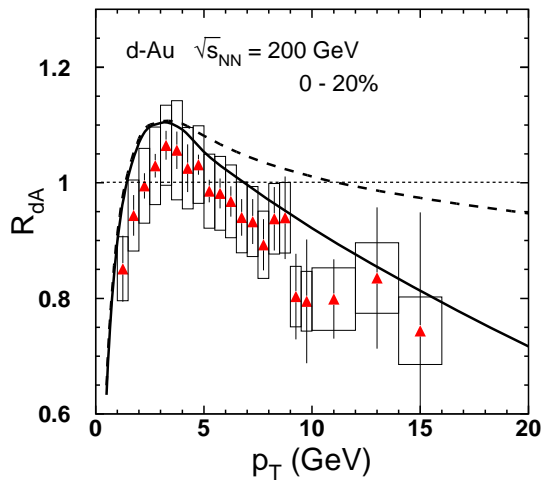}
\end{center} \caption{\label{fig:cronin-c}  (Color online) 
Nuclear attenuation factor $R_{dAu}(p_T)$ as function of
$p_T$ of  $\pi^0$ mesons produced in central ($0-20\%$) $d$-$Au$ collisions
at $\sqrt{s}=200\GeV$ and $\eta = 0$. The solid and dashed curves
represent the model predictions calculated with and without the ISI corrections.  Isotopic effect is included. The data are from the PHENIX experiment \cite{cronin-phenix}.
 }
 \end{figure}
There is a firm indication in data that the predicted strong nuclear suppression at large $x_T$ is indeed observed. Still, the statistical evidence of the effect needs to be enhanced.

In nuclear collisions the ISI effects are calculated similarly, using the modified 
parton distribution functions Eq.~(\ref{600}) for nucleons in both colliding nuclei.
The resulting complementary suppression reduces $R_{AA}(p_T)$ at large $x_T$.
This is demonstrated in Fig.~\ref{fig:rhic-0-5-gf}.
 \begin{figure}[htb]
\vspace*{5mm}
\centerline{
  \scalebox{0.4}{\includegraphics{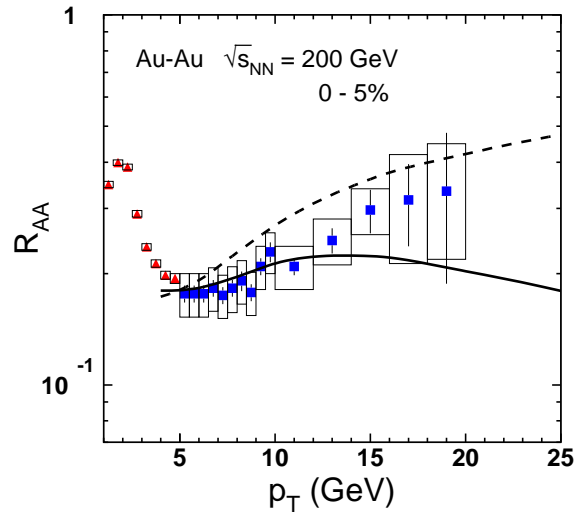}}}
\caption{\label{fig:rhic-0-5-gf} (Color online) Nuclear attenuation factor $R_{AA}(p_T)$ 
for neutral pions produced in central gold-gold collisions at
$\sqrt{s}$ = 200 $\GeV$.
The solid and dashed line are calculated with or without
ISI corrections.
PHENIX data are from \cite{phenix-b} (triangles)
and \cite{phenix-0} (squares).
}
 \end{figure}
If we just repeated the same calculations as done above for the LHC, we would get
$R_{AA}$ steeply rising with $p_T$, as is depicted by the dashed curve.
However the ISI effects and energy conservation impose a sizable additional suppression, as is shown by the solid curve. Apparently this improves the agreement with data.
Notice that the only fitted parameter, the transport coefficient, should be re-adjusted and was found to be  $\hat q_0=1.6\GeV^2\!/\!\fm$ at this energy, smaller than at $\sqrt{s}=2.76\TeV$, as expected.

Keeping this parameter fixed we can calculate other observables for gold-gold collisions at
$\sqrt{s}=200\GeV$. Fig.~\ref{fig:rhic200-b-gf} presents our results for the suppression of neutral pions at different collision centralities, in comparison with PHENIX data \cite{phenix-b}.
 \begin{figure}[htb]
\vspace*{5mm}
\centerline{
  \scalebox{0.4}{\includegraphics{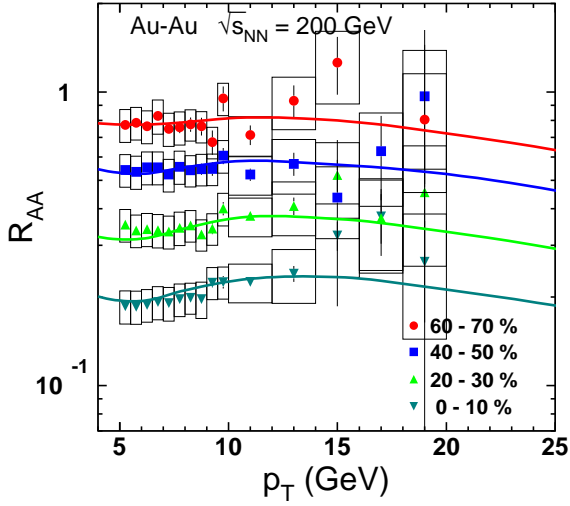}}}
\caption{\label{fig:rhic200-b-gf} (Color online) Centrality dependence of the suppression factor $R_{AA}(p_T,b)$ measured in the PHENIX experiment \cite{phenix-b} in gold-gold collisions at $\sqrt{s}=200\GeV$. The intervals of centrality are indicated in the plot.}
 \end{figure}\\


One can access even larger values of $x_T=2p_T/\sqrt{s}$, by either increasing $p_T$ or
going down in energy.  In both cases the effects of ISI should be stronger.
Indeed, data for $R_{AA}(p_T)$ in gold-gold collisions at $\sqrt{s}=62\GeV$ plotted in Fig.~\ref{fig:rhic62-b-gf} show a falling, rather than rising $p_T$-dependence.
 \begin{figure}[b]
\vspace*{5mm}
\centerline{
  \scalebox{0.4}{\includegraphics{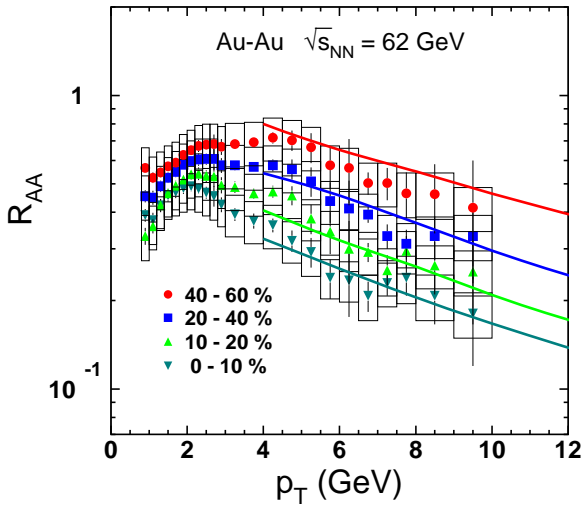}}}
\caption{\label{fig:rhic62-b-gf} (Color online) The same as in Fig.~\ref{fig:rhic200-b-gf},
but at $\sqrt{s}=62\GeV$. Data are from \cite{phenix62-b}.}
 \end{figure}
Our calculations using Eq.~(\ref{540}) again demonstrate good agreement. Of course,
the hot medium properties changed, so we had to re-adjust the parameter
$\hat q_0=1.2\GeV^2\!/\!\fm$.

\subsection{Azimuthal anisotropy}\label{v2}

Experimental observation of a suppression of high-$p_T$ hadrons escaping from the dense medium means that not the whole volume of the medium contributes.
Independently of the suppression mechanism (except for the contribution of ISI),
this means that the longer is the path-length of propagation in the medium (in two-dimensional transverse plane), the stronger is the suppression.
Thus, one can conclude that the direction of propagation normally to the medium surface  is preferable. For a non-central collision with an almond shape of the intersection area this should lead to an azimuthal asymmetry of high-$p_T$ hadron production.
Although this conclusion should be valid for any mechanism of suppression, 
the magnitude of the asymmetry is of course model dependent.

Our main result for high-$p_T$ hadron suppression, Eq.~(\ref{540}), can be tested by comparison of the predicted azimuthal angle dependence with data. Following data \cite{alice-phi-v2}, we split the integration over $\phi$ in (\ref{540}) into two intervals: (i) $|\phi|>3\pi/4$ plus $|\phi|<\pi/4$; (ii) $\pi/4<|\phi|<3\pi/4$. These two contributions are called in- and out-of-plane respectively. As expected, in-plane events are less suppressed compared with out-of-plane, as is depicted by the upper and bottom curves in Fig.~\ref{fig:lhc-phi-gf-all}.
 \begin{figure}[t]
\centerline{
  \scalebox{0.55}{\includegraphics{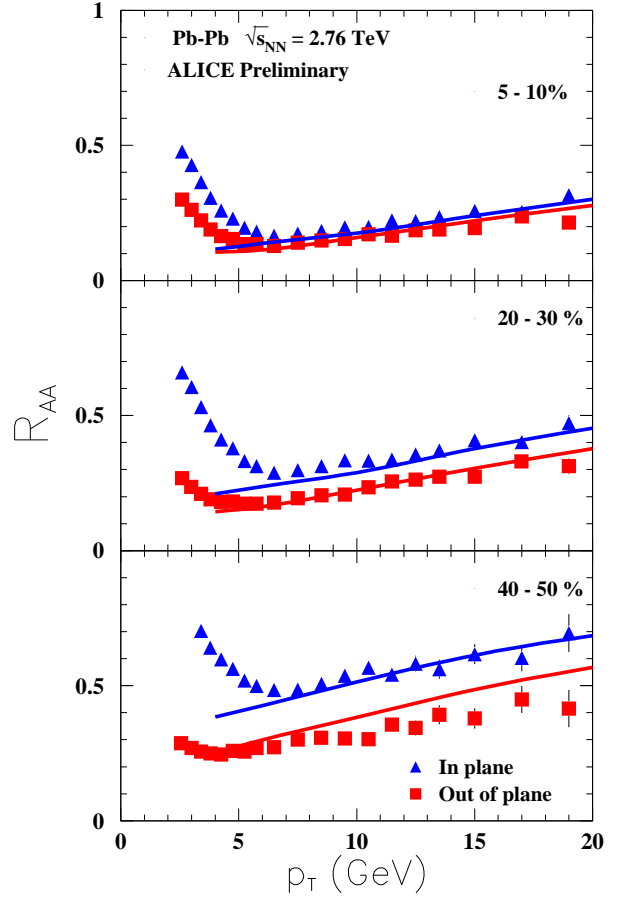}}}
\caption{\label{fig:lhc-phi-gf-all} (Color online) Nuclear factor $R_{AA}(p_T)$ for charge hadrons in lead-lead collisions at
$\sqrt{s}$ = 2.76 $\TeV$ at different centralities. ALICE data \cite{alice-phi-v2} and our calculations with Eq.~(\ref{540}) are divided into 
two classes: {\it In plane} ($3\pi/4<|\phi|<\pi$ and
$|\phi|<\pi/4$), and
{\it Out-of-plane} ($\pi/4<|\phi|<3\pi/4$).}
 \end{figure}
Our results agree well with ALICE data \cite{alice-phi-v2} at $p_T>6\GeV$ and at all measured centralities.

 \begin{figure}[htb]
\centerline{
  \scalebox{0.35}{\includegraphics{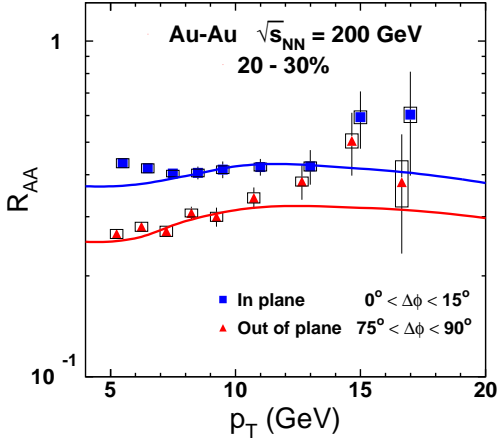}}}
\caption{\label{fig:phenix-phi} (Color online) $R_{AA}(p_T)$ for charge hadrons in gold-gold collisions at
$\sqrt{s}\eq 200\GeV$ at centralities $0-20\%$. PHENIX data \cite{phenix-b} and our calculations with Eq.~(\ref{540}) are divided into 
two classes of events: {\it In-plane} ($11\pi/12<|\phi|<\pi$ and
$|\phi|<\pi/12$), and
{\it Out-of-plane} ($5\pi/12<|\phi|<7\pi/12$).}
 \end{figure}

Similar measurements of $R_{AA}$ were performed in the PHENIX experiment for
the In- and Out-of the scattering plane events \cite{phenix-b}. The results for centrality 20-30\% are plotted in Fig.~\ref{fig:phenix-phi} in comparison with our calculations.
Notice that  both Figs.~\ref{fig:lhc-phi-gf-all} and \ref{fig:phenix-phi} show that the transition from the hydrodynamic to perturbative regimes occur for In-plane
events with a delay, at higher $p_T$. This is natural, since the hydrodynamic flow is much stronger, correspondingly the cross section is larger.

Usually data on azimuthal asymmetry of particle production are presented in terms of the 
second moment of the $\phi$-distribution, $v_2\equiv \left\la \cos(2\phi)\right\ra$.
We can calculate it with a slight modification of Eq.~(\ref{540}),
\begin{widetext}
\beq
v_2(p_T,b)=
\frac{\int d^2\tau \,T_A(\tau)T_B(\vec b-\vec\tau)\int\limits_0^{2\pi}d\phi
\cos(2\phi)\,
\left|\int\limits_0^\infty dr\,r\,\Psi_h(r) 
G_{\bar qq}(0,0;l_{max}, r)\right|^2}
{\int d^2\tau \,T_A(\tau)T_B(\vec b-\vec\tau)
\int\limits_0^{2\pi}d\phi 
\left|\int\limits_0^\infty dr\,r\,\Psi_h(r)\, 
G_{\bar qq}(0,0;l_{max}, r)\right|^2}.
\label{620}
\eeq
\end{widetext}
We remind that the Green function implicitly depends on the impact parameters $\vec b$ and $\vec\tau$, and on the trajectory of the dipole in the hot matter, and that these dependences are contained in the functions $\Omega(l_1,l_2)$ and $\Upsilon_i(l_1,l_2)$ in Eq.~(\ref{520}).

In order to compare with data on $v_2(p_T)$, one should integrate the numerator and denominator in (\ref{620}) over the intervals of impact parameter from $b_{min}$ to $b_{max}$, which correspond to  the measured intervals of centrality. Our results are compared with
ALICE data \cite{alice-phi-v2} and with CMS data \cite{cms-v2} in Figs.~\ref{fig:alice-v2-gf-all} and \ref{fig:cms-v2-gf-all} respectively. In all cases we observe good agreement. 
 \begin{figure}[htb]
\vspace*{5mm}
\centerline{
  \scalebox{0.45}{\includegraphics{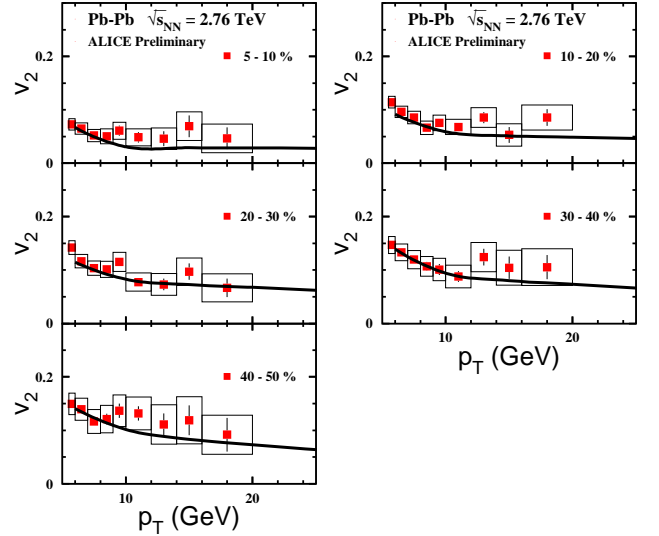}}}
\caption{\label{fig:alice-v2-gf-all} (Color online) ALICE data \cite{alice-phi-v2} for azimuthal anisotropy, $v_2$, vs $p_T$ for charge hadron
production in lead-lead collisions at mid rapidity, at
$\sqrt{s}$ = 2.76 $\TeV$ and at different
centralities indicated in the figure. The curves present the results of calculation with Eq.~(\ref{620}).
}
 \end{figure}

 \begin{figure}[htb]
\vspace*{5mm}
\centerline{
  \scalebox{0.45}{\includegraphics{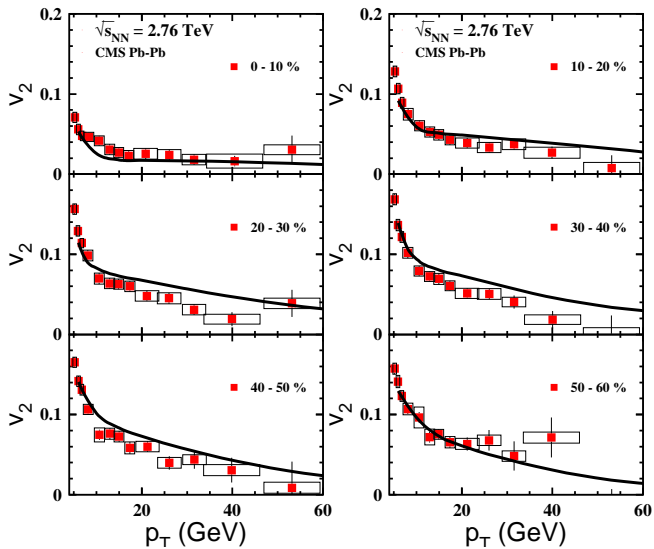}}}
\caption{\label{fig:cms-v2-gf-all} (Color online) The same as in Fig.~\ref{fig:alice-v2-gf-all}, but displaying data from the CMS experiment  \cite{cms-v2}.}
 \end{figure}

Naturally, as with data on $R_{AA}$, our pQCD calculations for $v_2(p_T)$ grossly underestimate data at small $p_T\lesssim6\GeV$. Remarkably, the transition to the pQCD regime occurs for $v_2(p_T)$ at the same $p_T$, as for $R_{AA}(p_T)$.
This confirms the presence of two different mechanisms: the dominant hydrodynamic mechanism of elliptic flow, providing a large and rising with $p_T$ anisotropy $v_2(p_T)$, which abruptly switches to the regime of pQCD, having a much smaller azimuthal anisotropy.
So it is not accidental that both $R_{AA}(p_T)$ and $v_2(p_T)$ swiftly change their behavior at the same value of $p_T$.

We also compare our results for the azimuthal anisotropy with RHIC data. Although we included the effects of ISI suppressing $R_{AA}$ at large $p_T$,
these corrections mostly cancel in the ratio, Eq.~(\ref{620}). Our calculations
agree well with PHENIX data for the azimuthal asymmetry for $\pi^0$ production at $\sqrt{s}=200\GeV$ and at mid rapidity, as is
demonstrated in Fig.~\ref{fig:rhic-v2-gf-all}.
 \begin{figure}[htb]
\vspace*{5mm}
\centerline{
  \scalebox{0.45}{\includegraphics{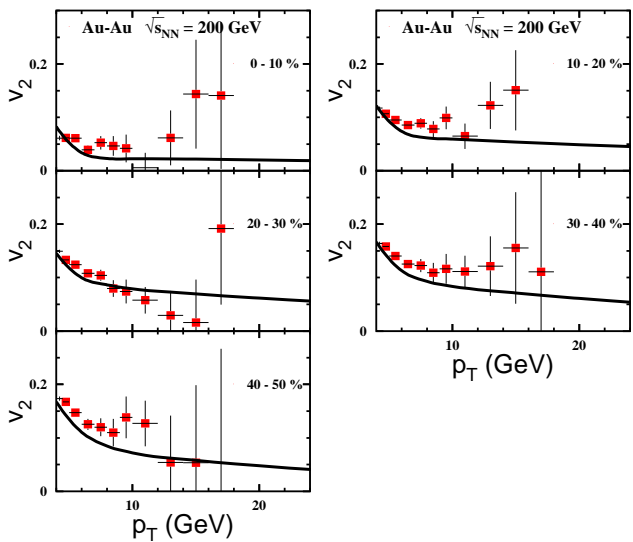}}}
\caption{\label{fig:rhic-v2-gf-all} (Color online) The same as in Fig.~\ref{fig:alice-v2-gf-all}, but displaying data from the PHENIX experiment \cite{phenix-v2}.
}
 \end{figure}

\section{Summary and prospectives}

This paper attempts at a quantitative understanding of experimentally observed strong
attenuation of hadrons inclusively produced with large transverse momenta in heavy ion collisions, based on contemporary models for space-time development of hadronization. 
The improved quality of data from RHIC and new high-statistics data from
LHC makes the test of models more challenging and also more decisive. 
In particular, the popular energy loss scenario, based on the unjustified assumption of
long production length, experiences difficulties explaining the new data. It also fails to explain the data from the HERMES experiment at HERA for leading hadron production in semi-inclusive DIS, which provides a sensitive testing ground for in-medium hadronization models.

Here we presented an alternative mechanism of suppression of high-$p_T$ hadrons inclusively produced in heavy ion collisions. First of all, we investigated the space-time development of gluon radiation and energy dissipation by a highly virtual parton
produced in a high-$p_T$ process. The key point of this consideration is an uniquely high initial virtuality of such a parton, which is of the order of its energy. Thus, increasing the jet energy one simultaneously enhances the hardness of the process, intensifying  the dissipation of the energy. For this reason energy conservation imposes tough constraints on the production length of leading hadrons, which does not rise with $p_T$, but remains short, as is demonstrated in Fig.~\ref{fig:mean-lp}.

It is worth emphasizing that the shortness of the production length is a specific feature of
inclusive hadron production. However, formation of a high-$p_T$ jet is characterized by a much longer time scale. Indeed, the convolution Eq.~(\ref{300}) of steeply falling  cross section of parton-parton scattering with parton distribution and fragmentation functions 
strongly enhances the contribution of large $z_h$, as is demonstrated in Fig.~\ref{fig:mean-zh}. In this case energy conservation is an important issue and restricts
the production length. As for fully reconstructed high-$p_T$ jets, they are characterized by a very different space-time development, and the medium-induced energy loss may be indeed an important source of the observed jet suppression.

With the production of a colorless hadronic state (dipole) the gluon radiation and energy loss cease, and the main reason for suppression of the production rate becomes the survival probability of the pre-hadron propagating through the dense matter. Apparently,
at larger $p_T$ the expansion of the initially small initial size of the dipole slows down due to Lorentz time dilation.
Here the effect of color transparency is at work, and the medium becomes more transparent for more energetic dipoles. 
A rising nuclear suppression factor $R_{AA}(p_T)$ is indeed observed at LHC,
in good accord with  the effect of color transparency.

In this paper we performed calculations based on the most strict quantum-mechanical 
description of space-time development and attenuation of color-dipoles propagating through a medium, the path-integral method. We calculated and compared with data the suppression factor $R_{AA}(b)$ for different centralities and energies of collision, ranging from $\sqrt{s}=62\GeV$ to $2.67\TeV$. Additional suppression arising from initial state multiple interactions, important at large $x_T$ or $x_L$, was also included. The related corrections are found to be important at $x_T\gsim0.1$, where they slow down the rise of $R_{AA}$ and even turn it into a falling $p_T$-dependence. 
We also calculated the azimuthal anisotropy
of hadron production, which reflects the asymmetric shape of the overlap area of the colliding nuclei. 

In all cases we reached good agreement with available data from RHIC and LHC at high $p_T$. The only adjusted parameter, the maximal value of the transport coefficient, Eq.~(\ref{340}), was found to be $\hat q_0=2\GeV^2\!/\!\fm$, $1.6\GeV^2\!/\!\fm$ and $1.2\GeV^2\!/\!\fm$ at $\sqrt{s}=2.76\TeV$, $200\GeV$ and $62\GeV$ respectively, for heavy nuclei, lead and gold.

Our results reproduce quite well data at $p_T\gsim 6\GeV$.
However, the observed $R_{AA}(p_T)$ and $v_2(p_T)$ expose quite  a different behavior towards smaller $p_T$,  steeply rising and shaping a bump. We relate this to an interplay of two mechanisms of hadron production: (i) evaporation of hadrons from the created hot medium controlled by hydrodynamics; (ii) perturbative QCD mechanism of high-$p_T$ production of hadrons, which propagate and attenuate in the hot medium.
The abrupt transition between the two mechanisms causes distinct minima in $R_{AA}(p_T)$ and in $v_2(p_T)$, both observed at the same values of $p_T$. We plan to work on combining the two mechanisms, aiming to describe data in the full range of $p_T$.

\section{Acknowledgments}

We are grateful to Hans-J\"urgen Pirner and Klaus Reygers for discussion of the results.
We thank Jacek Otwinowski for providing us with the updated results of the ALICE experiment.
This work was supported in part by Fondecyt (Chile) grants 1090236,
1090291 and 1100287. The work of B.Z.K. was
supported also by the Alliance Program of the Helmholtz Association, contract HA216/EMMI "Extremes of Density and Temperature: Cosmic Matter in the Laboratory".
The work of J.N. was partially supported by the
  Slovak Funding Agency, Grant 2/0092/10, by the Slovak Research and
  Development Agency APVV-0050-11 and by Grants VZ M\v SMT
  6840770039 and LA 08015 (Ministry of Education of the Czech Republic).

\end{document}